\documentclass[prb,aps,amsmath,amssymb,showpacs,twocolumn]{revtex4-1}

\usepackage{bm,color,float,dcolumn,amssymb,graphicx,hyperref,subfigure}

\newcommand{\overbar}[1]{\mkern 2.2mu\overline{\mkern-2.2mu#1\mkern-2.2mu}\mkern 2.2mu}

\begin{document}

\title{Topological phases of two-component bosons in \\ species-dependent artificial gauge potentials}

\author{Ying-Hai Wu$^{1,2}$ and Tao Shi$^{1}$}
\affiliation{$^1$ Max-Planck-Institut f{\"u}r Quantenoptik, Hans-Kopfermann-Stra{\ss}e 1, 85748 Garching, Germany \\
$^2$ Department of Physics, The Pennsylvania State University, University Park, PA 16802, USA}

\date{\today}

\begin{abstract}
We study bosonic atoms with two internal states in artificial gauge potentials whose strengths are different for the two components. A series of topological phases for such systems is proposed using the composite fermion theory and the parton construction. It is found in exact diagonalization that some of the proposed states may be realized for simple contact interaction between bosons. The ground states and low-energy excitations of these states are modeled using trial wave functions. The effective field theories for these states are also constructed and reveal some interesting properties.
\end{abstract}

\maketitle

\section{Introduction}

Atomic gases in the quantum degenerate regime provide unprecedented opportunities for studying quantum many-body systems \cite{Bloch1,Bloch2}. These charge neutral atoms do not couple to electromagnetic fields as charged particles do. To this end, various methods for generating artificial gauge potentials have been proposed and some of them have been successfully implemented \cite{Dalibard,Goldman1}. A conceptually simple way of generating an effective magnetic field for neutral atoms is using the Coriolis force in a rotating system \cite{Cooper1,Viefers}. There has also been substantial progress in generating Abelian and non-Abelian artificial gauge potentials using atom-light coupling in continuum or on optical lattices \cite{Lin1,Lin2,Aidelsburger1,Miyake,Jotzu,Aidelsburger2,Kennedy}. 

The synthesis of artificial gauge potentials extend the realm of phenomena that can be simulated using cold atoms to topological phases of matter. One primary example is the quantum Hall states \cite{Klitzing,Tsui} which occur in two-dimensional electron gases exposed to an external magnetic field (i.e. an Abelian gauge potential). Another example of interest in recent years is time reversal symmetric topological insulators \cite{Hasan,Qi} arising from spin-orbit couplings (which can be interpreted as a non-Abelian gauge potential). The cold atom systems allow us to study not only fermions but also bosons. The interaction potential between bosons can be modeled very well as the short-range contact interaction in most cases, but long-range interactions can be found in dipolar molecules \cite{Park}.

The topological phases of bosons in artificial gauge potentials have been studied in many previous works. For one-component bosons, composite fermion states and Moore-Read state can be realized using contact interaction \cite{Chang,Cooper2,Regnault1} while Read-Rezayi states may appear if long-range interactions are introduced \cite{Rezayi1,Cooper3}. For two-component bosons, composite fermion states can also be realized using contact interaction \cite{Wu1}. At filling factor $\nu=4/3$, a non-Abelian spin-singlet state has been studied as a possible candidate \cite{Grass1,Furukawa1} but there is a competing composite fermion state \cite{Wu1}. In addition to fractional quantum Hall (FQH) states, a symmetry protected topological (SPT) state without anyonic excitations can be realized at $\nu=2$ \cite{Wu1,Furukawa2,Regnault2,Grass2}. One interesting possibility for two-component bosons is creating artificial gauge potentials that depend on the atomic species; e.g., the direction of the potential for an atom is related to its magnetic moment \cite{Aidelsburger1,Miyake}. This is partially driven by the physics of two-dimensional topological insulators, for which a minimal model consisting of integer quantum Hall (IQH) states of spin-up electrons in positive magnetic field and spin-down electrons in negative magnetic field can be constructed. The quantum phases of two-component bosons in equal but opposite synthetic magnetic fields have been studied \cite{Furukawa3,Repellin}. In general, one may ask what could happen when multi-component bosons are coupled to general non-Abelian gauge potentials. In view of the successful identification of various topological phases in one- and two-component bosons, we can expect to see a plethora of exotic phenomena including FQH and SPT states due to the interplay between internal degrees of freedom, artificial gauge potentials, and tunable interactions. Theoretical investigations of such systems would be useful guidance for future experimental studies as they can help to narrow down the parameter space one should explore.

In this paper, we consider the cases where two types of atoms experience artificial gauge potentials pointing to the {\em same} direction but having {\em different} magnitudes. A straightforward way of creating such a system is to use a superposition of two potentials with different magnitudes: one has the same direction for both components and the other has opposite directions for the two components. The composite fermion theory \cite{Jain1} and parton construction \cite{Jain2,Jain3} will be used to construct trial wave functions describing topological phases in such systems. These trial wave functions are compared with exact diagonalization results. The physical properties of these states are also studied using effective field theories. In some cases, the systems can be interpreted as ``topological insulators" of emergent particles in the sense that the effective magnetic fields experienced by two types of composite fermions have the same magnitude but point to opposite directions.

\section{Models and Methods}

Although some experimental systems implement artificial gauge potentials in optical lattices, we shall consider continuum models for simplicity. This is a reasonable and convenient approximation because the Harper-Hofstadter model for particles moving in a periodic potential with magnetic fluxes \cite{Harper1,Wannier,Azbel,Hofstadter} reduces to a continuum model when the flux per plaquette is small enough \cite{Harper2}. For a system of bosons with two possible internal states (called spin-up and spin-down) in two dimensions, the single-particle Hamiltonian is
\begin{eqnarray}
H_{0} = \frac{1}{2M} \left[
\begin{array}{cc}
\left( {\mathbf p} - e^{\uparrow} {\mathbf A}^{\uparrow} \right)^{2} & 0 \\
0 & \left( {\mathbf p} - e^{\downarrow} {\mathbf A}^{\downarrow} \right)^{2}
\end{array} \right]
\label{SingleBody}
\end{eqnarray}
where $e^{\sigma}{\mathbf A}^{\sigma}$ is the artificial gauge potential for bosons with spin $\sigma$ and $M$ is their mass. The introduction of ``charge" $e^{\sigma}$ for the particles is redundant at this stage but will be useful later when we discuss the physical interpretation of our results. The physical quantities should only depend on the product $e^{\sigma}{\mathbf A}^{\sigma}$. The magnetic field strength is ${\nabla}\times{\mathbf A}^{\sigma}={\mathbf B}^{\sigma}=B^{\sigma}{\widehat e}_z$. We will focus on systems in which $e^{\uparrow}B^{\uparrow}$ and $e^{\downarrow}B^{\downarrow}$ have the same sign but different magnitudes ($e^{\downarrow}B^{\downarrow}$ is always taken to be smaller). The solution of the single-particle problem is two sets of Landau levels (LLs) for spin-up and spin-down bosons, respectively. In the symmetric gauge with ${\mathbf A}^{\sigma}=B^{\sigma}(-y/2,x/2,0)$, the lowest Landau level (LLL) wave functions are
\begin{eqnarray}
\psi^{\sigma}_{m}(z) = \frac{z^{m}\exp\left(-\frac{|z|^{2}}{4\ell^{2}_{\sigma}}\right)}{\ell^{m+1}_{\sigma}\sqrt{2{\pi}2^{m}{m!}}}
\label{SingleWave}
\end{eqnarray}
where $z=x+iy$ is the two-dimensional complex coordinate, $m$ is the $z$ component of the angular momentum, and $\ell_{\sigma}=\sqrt{{\hbar}/(e^{\sigma}B^{\sigma})}$ is the magnetic length. In later discussions, we will consider systems in which the particles carry negative charge or see a magnetic field along the $-{\widehat e}_{z}$ direction. Their wave functions can be obtained by taking the complex conjugate of Eq. \ref{SingleWave}.

For a many-body system with $N^{\sigma}_{b}$ bosons in the spin state $\sigma$, we perform exact diagonalization to find its ground state and low-lying excitations. The bosons are placed on a compact manifold such as sphere or torus \cite{Yoshioka,Haldane1,Haldane2} to eliminate edge effects. The number of magnetic fluxes through the surface of sphere or torus for the spin state $\sigma$ is denoted as $N^{\sigma}_{\phi}$. The filling factors $\nu^{\sigma}$ for the two spin states are defined separately because $N^{\uparrow}_{\phi}{\neq}N^{\downarrow}_{\phi}$. The magnetic lengths $\ell_{\uparrow,\downarrow}$ for the two spin states are different and we choose the smaller one $\ell_{\uparrow}$ as the length scale. The two-body contact interaction potential between the bosons is
\begin{eqnarray}
V({\mathbf r}_{i}-{\mathbf r}_{j}) = \delta({\mathbf r}_{i}-{\mathbf r}_{j}) 
\left(\begin{array}{cc}
g_{\uparrow\uparrow}   &  g_{\uparrow\downarrow} \\
g_{\downarrow\uparrow} &  g_{\downarrow\downarrow} 
\end{array}\right)
\label{ManyBody}
\end{eqnarray}
where $g_{\sigma\tau}$'s characterize the interaction strengths. In the pseudopotential representation of this interaction \cite{Haldane1}, the only non-vanishing component is the zeroth one $V_{0}$. We choose $g_{\sigma\tau}=g=4{\pi}\ell^{2}_{\uparrow}$ such that $V_{0}=1$ in the spin-up wave function basis and it is used as the unit of energy in what follows. It is assumed that the bosons are confined to their respective LLLs and mixing with higher Landau levels is negligible. In the second quantized notation, the many-body Hamiltonian is
\begin{eqnarray}
\frac{1}{2} \sum_{\sigma\tau} \sum_{\{m_{i}\}} g^{\sigma\tau} F^{\sigma\tau\tau\sigma}_{m_{1}m_{2}m_{4}m_{3}} C^\dagger_{\sigma,m_{1}} C^\dagger_{\tau,m_{2}} C_{\tau,m_{4}} C_{\sigma,m_{3}}
\label{ManyHamilton}
\end{eqnarray}
where the coefficients $F^{\sigma\tau\tau\sigma}_{m_{1}m_{2}m_{4}m_{3}}$ can be calculated using the wave functions (see Appendix A for details). 

\subsection{Sphere}

For the spherical geometry, a radial magnetic field can be generated by a magnetic monopole at the center of the sphere. The LLL wave functions on the sphere are
\begin{eqnarray}
\psi^{N^{\sigma}_{\phi}}_{m} = \left[ \frac{N^{\sigma}_{\phi}+1}{4\pi} \binom{N^{\sigma}_{\phi}}{N^{\sigma}_{\phi}-m} \right]^{\frac{1}{2}} u^{N^{\sigma}_{\phi}/2+m} v^{N^{\sigma}_{\phi}/2-m}
\end{eqnarray}
where $u=\cos(\theta/2)e^{i\xi/2},v=\sin(\theta/2)e^{-i\xi/2}$ are spinor coordinates ($\theta$ and $\xi$ are the azimuthal and radial angles in the spherical coordinate system) and $m$ is the $z$ component of the angular momentum. The radius $R$ of the sphere is $\ell_{\uparrow}\sqrt{N^{\uparrow}_\phi/2}=\ell_{\downarrow}\sqrt{N^{\downarrow}_\phi/2}$. The matrix element $F^{\sigma\tau\tau\sigma}_{m_{1}m_{2}m_{4}m_{3}}\propto\delta_{m_{1}+m_{2},m_{3}+m_{4}}$ where $\delta_{i,j}$ is the usual Kronecker delta function. This means that the $z$ component of the total angular momentum is a conserved quantity. The many-body eigenstates can be further classified by their total angular momentum eigenvalue $L(L+1)$. 

\subsection{Torus}

For the torus geometry, we consider a rectangle torus spanned by the basis vectors ${\mathbf L}_1=L_1{\widehat e}_x$ and ${\mathbf L}_2=L_2{\widehat e}_y$ and choose the vector potential ${\mathbf A}^{\sigma}=(0,B^{\sigma}x,0)$. The LLL wave functions on the torus are
\begin{eqnarray}
\psi^{N^{\sigma}_{\phi}}_m &=& \frac{1}{(\sqrt{\pi}L_{2}\ell_{\sigma})^{1/2}} \sum^{\mathbb{Z}}_{k} \exp \Big\{ i \frac{2{\pi}y}{L_{2}} \left( m+kN^{\sigma}_{\phi} \right) \nonumber \\
&-& \frac{1}{2} \left[ \frac{x}{\ell_{B}} - \frac{2{\pi}\ell_B}{L_{2}} \left( m+kN^{\sigma}_{\phi} \right) \right]^{2}  \Big\}
\end{eqnarray}
where the magnetic length $\ell_{\sigma}=\sqrt{L_{1}L_{2}/(2{\pi}N^{\sigma}_{\phi})}$. The matrix element $F^{\sigma\tau\tau\sigma}_{m_{1}m_{2}m_{4}m_{3}}\propto{\widetilde\delta}^{N^{\rm G}_{\phi}}_{m_{1}+m_{2},m_{3}+m_{4}}$ where ${\widetilde\delta}^{N^{\rm G}_{\phi}}_{i,j}$ is a generalized Kronecker delta function defined as
\begin{eqnarray}
{\widetilde\delta}^{N^{\rm G}_{\phi}}_{i,j}=1 \;\; {\rm iff} \;\; i \; {\rm mod} \; N^{\rm G}_{\phi} = j \; {\rm mod} \; N^{\rm G}_{\phi}
\end{eqnarray}
with $N^{\rm G}_{\phi}$ being the greatest common divisor of $N^{\uparrow}_{\phi}$ and $N^{\downarrow}_{\phi}$. This means that the many-body eigenstates can be labeled by a special momentum quantum number $K \equiv (\sum_{\sigma=\uparrow,\downarrow}\sum^{N^{\sigma}_{b}}_{i=1} m^{\sigma}_{i}) \; {\rm mod} \; N^{\rm G}_{\phi}$.

\section{Trial Wave Functions}

As the parameter space of our problem is quite large, we turn to the composite fermion theory \cite{Jain1} for guidance about where to search for gapped topological phases. This theory was originally developed for electronic systems, where one electron binds an even number of magnetic flux quanta (one flux quantum is $\phi_0=hc/e$) to become one composite fermion. The composite fermions move in effective magnetic field and form IQH states in many situations because they generally interact weakly with each other. For bosons in artificial gauge potentials, it is not immediately clear how to define ``flux quantum" as they do not carry charges. The simple choice used in previous works is to take the charge of the bosons to be 1 and attach an odd number of magnetic flux quanta to each boson. For one-component bosons and two-component bosons in the same magnetic fluxes, this prescription yields accurate approximations to the exact eigenstates of contact interaction \cite{Chang,Wu1}. If this is still applied when the two components experience different artificial gauge potentials, the magnetic field strengths for spin-up and spin-down bosons would be different. For certain system parameters, we will have cases in which the effective magnetic fields for spin-up and spin-down composite fermions point to opposite directions. This can not happen if the two components experience the same magnetic fluxes as studied before \cite{Wu1}.

The heuristic picture used above is captured by the trial wave functions
\begin{eqnarray}
\Psi_{[f_{\uparrow},f_{\downarrow}]} \sim \chi^{\uparrow}_{f_{\uparrow}} \left[ \chi^{\downarrow}_{f_{\downarrow}} \right]^* \chi_{1}
\label{ManyWave}
\end{eqnarray}
where $\chi_{f}$ is the wave function of $f$ completely filled LLs (without the Gaussian factor). The $\chi^{\uparrow}_{f_{\uparrow}}$ ($[\chi^{\downarrow}_{f_{\downarrow}}]^*$) factor only contains the coordinates of spin-up (spin-down) bosons, which describes spin-up (spin-down) composite fermions in positive (negative) effective magnetic field. The factor $\chi_1=\prod_{j<k}(z_j-z_k)$ contains the coordinates of bosons in both spin states and implements vortex attachment for all of them. The reason for using a sim sign rather than an equal sign is two fold. First of all, we do not include the Gaussian factors of the wave functions $\chi_{f}$ when constructing Eq. \ref{ManyWave} but only supplement the final result with the correct Gaussian factor. In addition, the wave function should be projected to the LLL for comparison with exact diagonalization results. For the state described by Eq. \ref{ManyWave} with $N^{\uparrow}_{b}$ spin-up bosons and $N^{\downarrow}_{b}$ spin-down bosons, the two spin states have filling factors $\nu^{\uparrow}=N^{\uparrow}_{b}/[N^{\uparrow}_{b}(f_{\uparrow}+1)/f_{\uparrow}+N^{\downarrow}_{b}]$ and $\nu^{\downarrow}=N^{\downarrow}_{b}/[N^{\downarrow}_{b}(f_{\downarrow}-1)/f_{\downarrow}+N^{\uparrow}_{b}]$.

An alternative understanding of the trial wave functions can be obtained using the parton construction \cite{Jain2,Jain3}. In this framework, one physical boson is decomposed to multiple fictitious partons, which move in the same magnetic field as the physical bosons and form IQH states. The state of the physical bosons is obtained by gluing the partons together. When the partons form IQH states, we may obtain gapped topological phases of the physical bosons. This interpretation requires $B^{\uparrow}=B^{\downarrow}$ so we have to choose $e^{\uparrow}$ and $e^{\downarrow}$ to be different. For the trial wave function Eq. \ref{ManyWave}, one physical boson with spin $\sigma$ consists of one type ${\rm I}$ parton with spin $\sigma$ and one spinless type ${\rm II}$ parton. The spin-up (spin-down) type ${\rm I}$ parton form the $\chi^\uparrow_{f_{\uparrow}}$ ($[\chi^\downarrow_{f_{\downarrow}}]^*$) state and the spinless type ${\rm II}$ parton forms the $\chi_1$ state. 

\section{Numerical Results}

\begin{figure}
\includegraphics[width=0.45\textwidth]{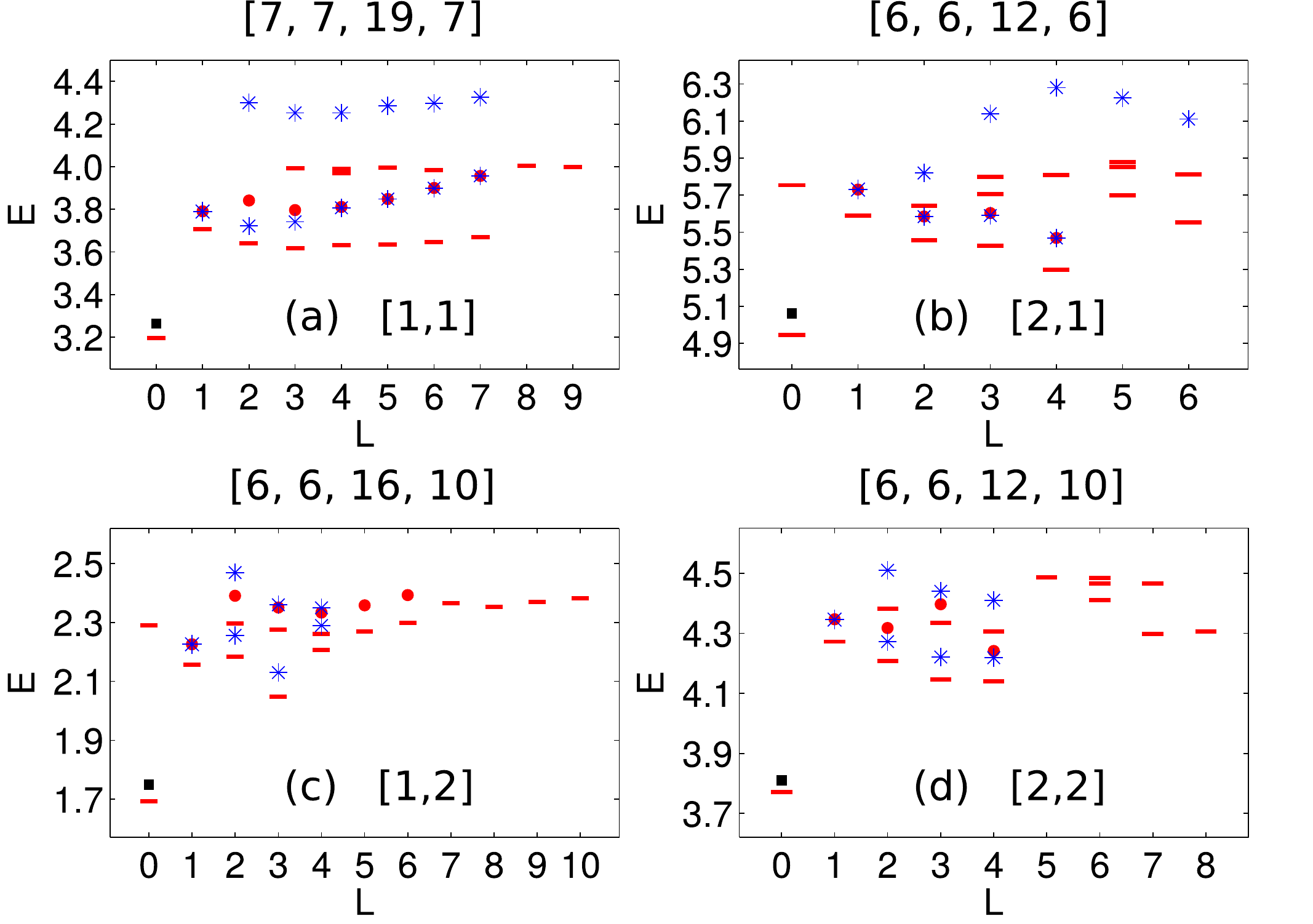}
\caption{Energy spectra of bosons on the sphere. The system parameters are given as $[N^{\uparrow}_{b},N^{\downarrow}_{b},N^{\uparrow}_{\phi},N^{\downarrow}_{\phi}]$ on top of each panel and as $[f_{\uparrow},f_{\downarrow}]$ inside each panel. The dots and asterisks represent the energies of the trial wave functions Eq.~(\ref{ManyWave}). In some cases, these two types of symbols have very close energy values and may not be easy to discern by inspection.}
\label{Figure1}
\end{figure}

\begin{figure}
\includegraphics[width=0.45\textwidth]{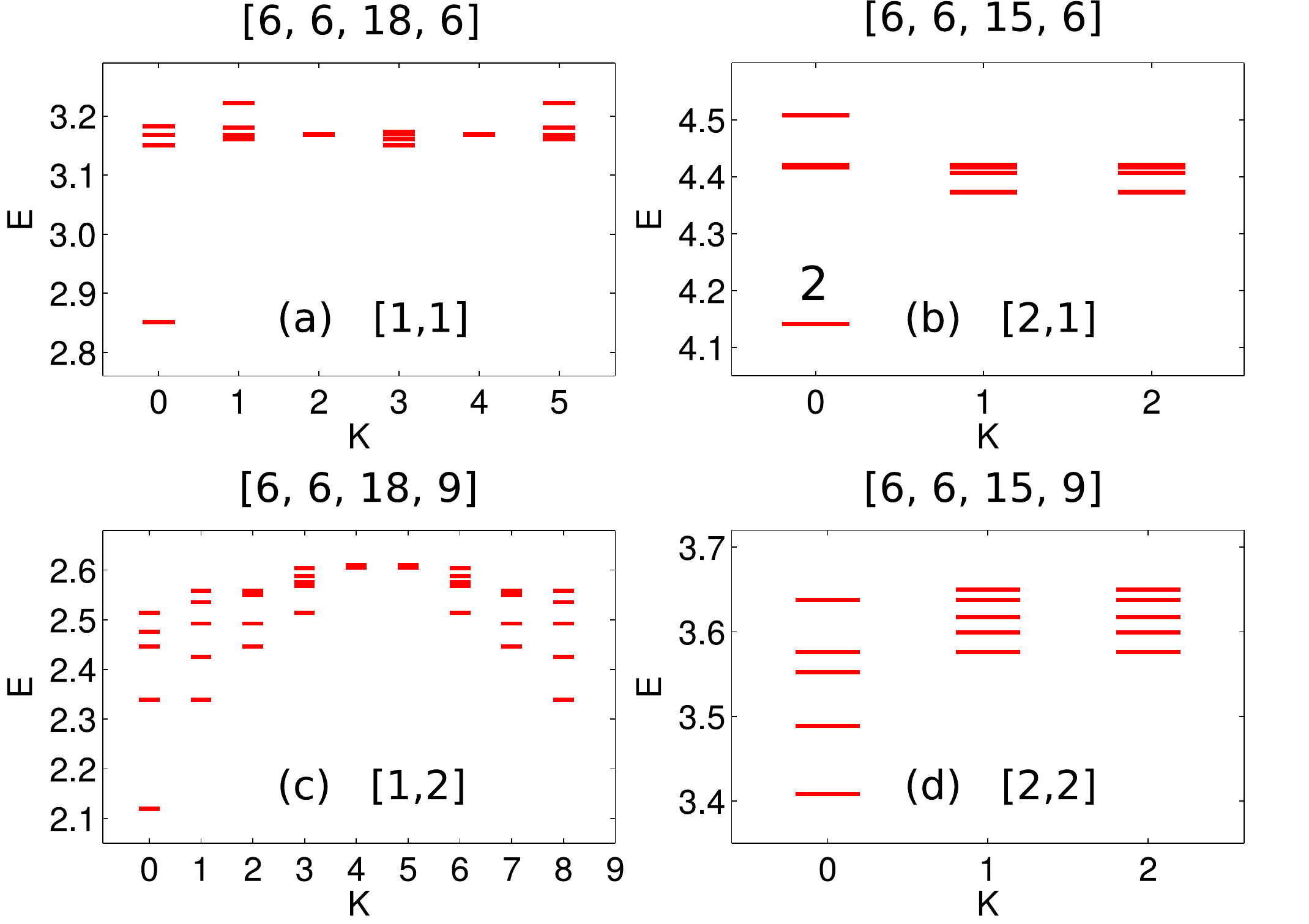}
\caption{Energy spectra of bosons on the torus. The aspect ratio $L_{1}/L_{2}$ is one for all panels. The system parameters are given as $[N^{\uparrow}_{b},N^{\downarrow}_{b},N^{\uparrow}_{\phi},N^{\downarrow}_{\phi}]$ on top of each panel and as $[f_{\uparrow},f_{\downarrow}]$ inside each panel. The number $2$ in panel (b) indicates two quasi-degenerate states which can not be resolved by eye.}
\label{Figure2}
\end{figure}

\begin{figure}
\includegraphics[width=0.45\textwidth]{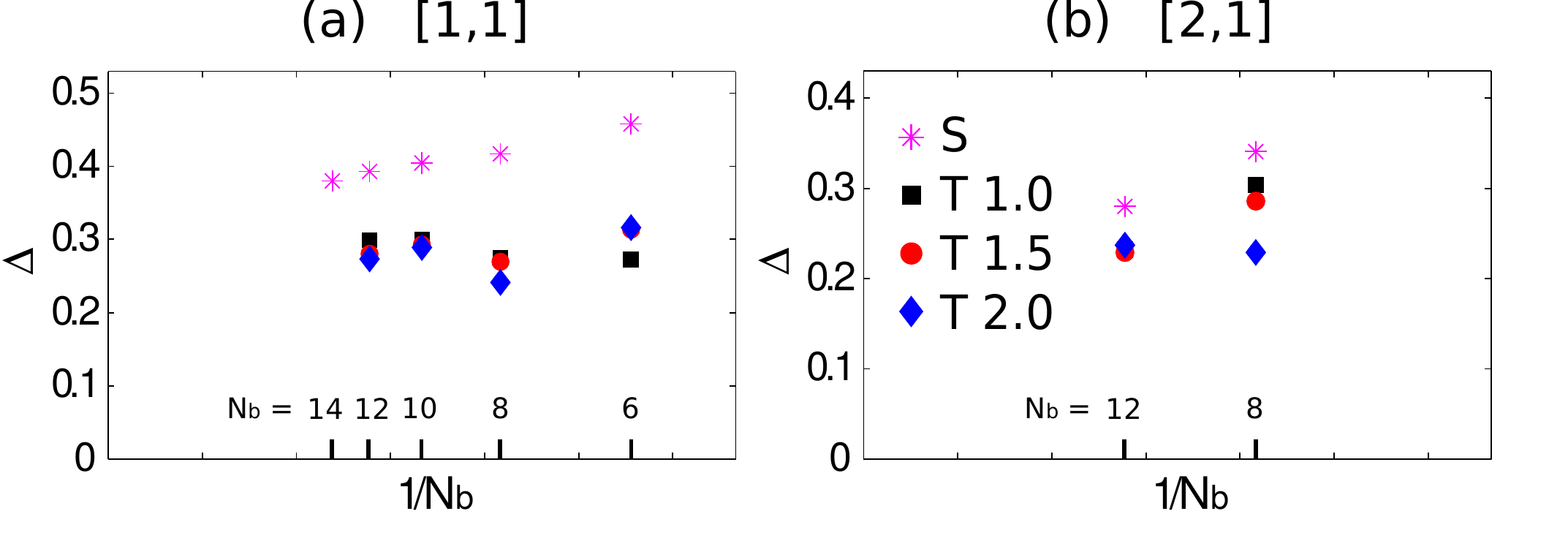}
\caption{Energy gap values on the sphere and the torus. The horizontal axis is the inverse of the total number of bosons $N_{b}=N^{\uparrow}_{b}+N^{\downarrow}_{b}$. The values on the sphere have been rescaled by a factor $\nu^{\uparrow}N^\uparrow_{\phi}/N^{\uparrow}_{b}$ to account for the deviation of the magnetic length \cite{Morf}. (a) The $f_{\uparrow}=1,f_{\downarrow}=1$ state; (b) the $f_{\uparrow}=2,f_{\downarrow}=1$ state. The legend is shown in panel (b) where ``S" means sphere, ``T" means torus, and the number is the aspect ratio $L_{1}/L_{2}$ of the torus.}
\label{Figure3}
\end{figure}

\begin{table*}
\centering
\begin{tabular}{ccccccccc}
\hline\hline
       &       &       &       &  $L$  &       &       &       &       \\\hline
Figure &    0  &   1   &   2   &   3   &   4   &   5   &   6   &   7   \\\hline
 1(a)  & 0.973 & 0.959 & 0.858 & 0.887 & 0.880 & 0.842 & 0.811 & 0.785 \\
       & (60164) & (179968) & (298507) & (414810) & (528122) & (637534) & (742421) & (841924) \\
 1(b)  & 0.949 & 0.929 & 0.932 & 0.901 & 0.907 &   -   &   -   &   -   \\
       & (1717) & (4810) & (8118) & (11007) & (13989) &  &  &  \\
 1(c)  & 0.965 & 0.954 & 0.937/0.686 & 0.941/0.935 & 0.922/0.929 & 0.931 & 0.919 &   -   \\
       & (25170) & (73991) & (123317) & (170210) & (216629) & (259711) & (301516) & \\
 1(d)  & 0.985 & 0.962 & 0.956/0.909 & 0.962/0.917 & 0.949/0.919 &   -   &   -   &   -   \\
       & (9951) & (28960) & (48312) & (66303) & (84117) &  &  &  \\
\hline\hline
\end{tabular}
\caption{The overlaps between trial wave functions and exact eigenstates shown in Fig. \ref{Figure1}. $L$ is the total angular momentum quantum number. The total number of linearly independent $L$ multiplets is given in parentheses below each overlap. The entries with one (two) number(s) are the overlaps between the trial wave functions represented by the dots (asterisks) and the exact states.}
\label{Table1}
\end{table*}

We show in Fig. \ref{Figure1} and Fig. \ref{Figure2} the energy spectra of bosons on sphere and torus with four difference choices of $f_{\uparrow}$ and $f_{\downarrow}$ in Eq. \ref{ManyWave}. For bosons on the sphere, we are able to construct the trial wave functions explicitly. The overlaps between the trial states and the exact eigenstates are shown in Table \ref{Table1}. The trial wave functions for the ground states (black squares in Fig. \ref{Figure1}) are excellent approximations. The trial wave functions for the excited states can be obtained by promoting one composite fermion to the lowest empty composite fermion LL. For two component bosons in the same artificial gauge potentials \cite{Wu1}, one should consider all possible ways of exciting one composite fermion to account for the low-lying excitations. The same construction in our system results in the blue asterisks in Fig. \ref{Figure1} (i.e. the states with one excited spin-up {\em or} spin-down composite fermion). The asterisks are accurate approximations of the low-lying excitations in Fig. \ref{Figure1} (c) and (d), but some of them have quite high energies (so do not represent the low-lying excitations) in Fig. \ref{Figure1} (a) and (b). To further reveal the difference between our systems and those studied before, we construct the trial wave functions with {\em only} one spin-up composite fermion being excited and give their energies as red dots in Fig. \ref{Figure1}, which turn out to be a reasonably good description of the low-lying excitations in Fig. \ref{Figure1} (a) and (b) but not in Fig. \ref{Figure1} (c) and (d). In some angular momentum sectors, these two methods result in the same states so the dots and asterisks coincide with each other. The fact that the neutral excitations of the $f_{\uparrow}=1,f_{\downarrow}=1$ and the $f_{\uparrow}=2,f_{\downarrow}=1$ states can be modeled very well using one type of composite fermions makes them very different from those studied in previous works \cite{Wu1}. The energy gap values for several different systems corresponding to the $f_{\uparrow}=1,f_{\downarrow}=1$ and the $f_{\uparrow}=2,f_{\downarrow}=1$ states are plotted in Fig. \ref{Figure3}. Although it is difficult to perform an accurate finite size scaling analysis, it is plausible that these two states are gapped in the thermodynamic limit. We also note that finite size effects are stronger on the sphere because the states appear to have larger gaps on the sphere. 

It is useful to consider the limit where $f_{\uparrow}$ and/or $f_{\downarrow}$ become very large. For the case with both $f_{\uparrow,\downarrow}\rightarrow\infty$, the trial wave function Eq. \ref{ManyWave} actually describes composite fermions in zero effective magnetic field that form a gapless Fermi liquid whose low-lying excitations contain excited spin-up and spin-down composite fermions \cite{Wu2}. If we choose one of $f_{\uparrow,\downarrow}$ to be small and the other to be large, it is very likely that the system would also be gapless due to the composite fermions with large effective filling factor. We expect that Fermi-liquid-like features will show up in a system with sufficiently large $f_{\uparrow}$ and/or $f_{\downarrow}$. The precise values at which the system ceases to be gapped can not be determined for two reasons: (1) the gap value for a specific set of $f_{\uparrow}$ and $f_{\downarrow}$ should be extracted from finite size scaling but there are not enough available data points in many cases; (2) the systems with large $f_{\uparrow}$ and/or $f_{\downarrow}$ can only be defined for a large number of bosons so can not be studied using exact diagonalization. We note that the energy spectrum of the $f_{\uparrow}=2,f_{\downarrow}=2$ state on the torus [Fig. \ref{Figure2} (d)] has a small energy gap. This is further supported by effective field theory analysis presented below. It is possible that the $f_{\uparrow}=2,f_{\downarrow}=2$ state may actually be gapless in the thermodynamic limit. The fact that its low-lying excitations are better modeled using composite fermions of both types is also consistent with the behavior in the limit of infinite $f_{\uparrow}$ and $f_{\downarrow}$. 

However, the analysis presented above does not seem to be applicable to the $f_{\uparrow}=1,f_{\downarrow}=2$ state. We can not claim that the system is gapless in the thermodynamic limit based on the numerical evidence. On the other hand, its low-lying excitations are different from those of the probably gapped states at $f_{\uparrow}=1,f_{\downarrow}=1$ and $f_{\uparrow}=2,f_{\downarrow}=1$, which suggests that a different physical interpretation for this state is required. The effective field theories presented below provide some hints about how to resolve this issue.

\begin{widetext}

\section{Effective Field Theories}

We construct effective field theories for our systems in two ways. In the first approach, we take the charges of all bosons to be unit and use the Chern-Simons transformation to implement flux attachment \cite{Lopez1,Lopez2,Halperin}. The partition function of the system is ${\mathcal Z}=\int D[\psi^*,\psi] \exp(i \int dt \; d^2 {\mathbf r} \; \mathcal{L})$ with Lagrangian density
\begin{eqnarray}
\mathcal{L} = \sum_{\sigma=\uparrow\downarrow} \psi^{*}_{\sigma}(t,{\mathbf r}) \left[ i\partial_t - \frac{1}{2M} \left( {\mathbf p} - {\mathbf A}^{\sigma} \right)^2 \right] \psi_{\sigma}(t,{\mathbf r}) - \frac{1}{2} \sum_{\sigma,\tau=\uparrow\downarrow} \int d^2 {\mathbf r}^{\prime} \rho_{\sigma}(t,{\mathbf r}) V_{\sigma\tau}({\mathbf r}-{\mathbf r}^{\prime}) \rho_{\tau}(t,{\mathbf r}^{\prime})
\label{RealBoseAction}
\end{eqnarray}
where $\rho_{\sigma}(t,{\mathbf r})=\psi^{*}_{\sigma}(t,{\mathbf r})\psi_{\sigma}(t,{\mathbf r})$ is the density of bosons with spin $\sigma$. We define ${\widetilde\psi}({\mathbf r})=\psi({\mathbf r})\exp\left[-i\int d^2 {\mathbf r}^{\prime} \; {\rm arg}({\mathbf r}-{\mathbf r}^{\prime}) \rho({\mathbf r})\right]$ with ${\rm arg}({\mathbf r}-{\mathbf r}^{\prime})$ being the angle between the vector ${\mathbf r}-{\mathbf r}^{\prime}$ and the $x$ axis. The Lagrangian density is transformed to
\begin{eqnarray}
{\mathcal L} &=& \sum_{\sigma=\uparrow,\downarrow} {\widetilde\psi}^{*}_{\sigma}(t,{\mathbf r}) \left[ \left( i\partial_{t} - a_{0} \right) - \frac{1}{2M} \left( {\mathbf p} - {\mathbf A}^{\sigma} + {\mathbf a} \right)^2 \right] {\widetilde\psi}_{\sigma}(t,{\mathbf r}) \nonumber \\
&+& \frac{1}{4\pi} \epsilon_{\lambda\mu\nu} a_{\lambda}(t,{\mathbf r}) \partial_{\mu} a_{\nu}(t,{\mathbf r}) - \frac{1}{2} \sum_{\sigma,\tau=\uparrow\downarrow} \int d^2 {\mathbf r}^{\prime} \; {\widetilde\rho}_{\sigma}(t,{\mathbf r}) V_{\sigma\tau}({\mathbf r}-{\mathbf r}^{\prime}) {\widetilde\rho}_{\tau}(t,{\mathbf r}^{\prime})
\label{CompFermiAction}
\end{eqnarray}
where ${\widetilde\rho}_{\sigma}(t,{\mathbf r})={\widetilde\psi}^{*}_{\sigma}(t,{\mathbf r}){\widetilde\psi}_{\sigma}(t,{\mathbf r})$, the Chern-Simons vector potential ${\mathbf a}(t,{\mathbf r})=\int d^2 {\mathbf r}^{\prime} \; \rho(t,{\mathbf r}^{\prime}) [{\widehat z}\times({\mathbf r}-{\mathbf r}^{\prime})]/|{\mathbf r}-{\mathbf r}^{\prime}|^2$, and the Chern-Simons scalar potential $a_0$ is a Lagrangian multiplier. 

In the first line of Eq. \ref{CompFermiAction}, we introduce a weak electromagnetic potential $A^{\sigma}_{{\rm ext},\mu}$ in addition to ${\mathbf A}^{\sigma}$ to obtain
\begin{eqnarray}
{\mathcal L}_{\rm CF} = \sum_{\sigma=\uparrow,\downarrow} {\widetilde\psi}^{*}_{\sigma}(t,{\mathbf r}) \left[ (i\partial_{t} - a_{0} + A^{\sigma}_{{\rm ext},0} ) - \frac{1}{2M} \left( {\mathbf p} - {\mathbf A}^{\sigma} + {\mathbf a} - {\mathbf A}^{\sigma}_{{\rm ext}} \right)^2 \right] {\widetilde\psi}_{\sigma}(t,{\mathbf r})
\end{eqnarray}
In the second line of Eq. \ref{CompFermiAction}, the interaction term can be written as $g{\widetilde\rho}(t,{\mathbf r})\delta({\mathbf r}-{\mathbf r}^{\prime}){\widetilde\rho}(t,{\mathbf r}^{\prime})$ with ${\widetilde\rho}={\widetilde\rho}_{\uparrow}+{\widetilde\rho}_{\downarrow}=(2\pi)^{-1}\nabla\times{\mathbf a}$ so we have
\begin{eqnarray}
{\mathcal L}_{\rm CS} = \frac{1}{4\pi} \epsilon_{\lambda\mu\nu} a_{\lambda}(t,{\mathbf r}) \partial_{\mu} a_{\nu}(t,{\mathbf r}) - \frac{g}{8\pi^2} \int d^2 {\mathbf r}^{\prime} \; \left[ \nabla\times{\mathbf a}(t,{\mathbf r}) \right] \delta({\mathbf r}-{\mathbf r}^{\prime}) \left[ \nabla\times{\mathbf a}(t,{\mathbf r}^{\prime}) \right] \nonumber
\end{eqnarray}
The current $J^{\sigma}_{\mu}$ is related to $A^{\tau}_{{\rm ext},\nu}$ via the linear response equation $J^{\sigma}_{\mu}=K^{\sigma\tau}_{\mu\nu}A^{\tau}_{{\rm ext},\nu}$ and the magnetoplasmon excitations are related to the poles of the response function $K^{\sigma\tau}_{\mu\nu}$. The partition function can be written as $Z[A^{\sigma}_{\rm ext}] = \int D[a] \exp(iS_{\rm CS}) \int D[{\widetilde\psi}^*,{\widetilde\psi}] \exp(iS_{\rm CF})$ with $S_{\rm CF}=\int d^3 r \; {\mathcal L}_{\rm CF}$ and $S_{\rm CS}=\int d^3 r \; {\mathcal L}_{CS}$. The first step is to integrate out the composite fermion field ${\widetilde\psi}^*$ and ${\widetilde\psi}$ to obtain $Z[A^{\sigma}_{\rm ext}] = \int D[a] \exp(iS_{\rm eff})$. The second step is to integrate out the Chern-Simons gauge field in $S_{\rm eff}$ to obtain $Z[A^{\sigma}_{\rm ext}]=\exp \left( iS_{\rm ext} \right)$. The gap of the magnetoplasmon mode at zero momentum is (see Appendix B for more details)
\begin{eqnarray}
\Delta_0 = \frac{1}{2} \left[ \sqrt{ (f_{\uparrow}+1)^{2}{\overbar\omega}^{2}_{\uparrow} + (f_{\downarrow}-1)^{2}{\overbar\omega}^{2}_{\downarrow} + 2(f_{\uparrow}f_{\downarrow}+f_{\uparrow}-f_{\downarrow}+1){\overbar\omega}_{\uparrow}{\overbar\omega}_{\downarrow} } - (f_{\uparrow}+1){\overbar\omega}_{\uparrow} - (f_{\downarrow}-1){\overbar\omega}_{\downarrow} \right]
\label{GapValue}
\end{eqnarray}
where ${\overbar\omega}_{\sigma}$ is the effective cyclotron frequency of composite fermions. It should be emphasized that the gap values obtained in exact diagonalization generally are not at zero momentum. The finite momentum gap values are usually smaller than the zero momentum ones so the latter still provides useful information about whether the system is gapped. For the case with $f_{\uparrow}=f_{\downarrow}$, Eq. \ref{GapValue} simplifies to $\Delta_0=(\sqrt{f^{2}_{\uparrow}+1}-f_{\uparrow}){\overbar\omega}$ which suggests that the gap descreases as $f_{\uparrow}$ increases. This is consistent with the numerical result that the gap at $f_{\uparrow}=1,f_{\downarrow}=1$ is more robust than the gap at $f_{\uparrow}=2,f_{\downarrow}=2$. Another important observation is that the gap at $f_{\uparrow}=1,f_{\downarrow}=2$ vanishes, which suggests that the gap as seen in numerical calculations may be due to strong finite size effects. In general, Eq. \ref{GapValue} suggests that the energy gaps for all the states at $f_{\downarrow}=f_{\uparrow}+1$ are zero.

The ground state degeneracy of the many-body states on the torus can also be computed using the flux attachment approach together with the functional bosonization method \cite{Chen}, but we defer this calculation to Appendix B because we find that it can be obtained more easily using the parton construction \cite{Wen}. The spin-up type ${\rm I}$ parton, spin-down type ${\rm I}$ parton, and spinless type ${\rm II}$ parton have charges $e_{{\rm I}\uparrow}=f_{\downarrow}/(2f_{\uparrow}f_{\downarrow}-f_{\uparrow})$, $e_{{\rm I}\downarrow}=-1/(2f_{\downarrow}-1)$, and $e_{{\rm II}}=2f_{\downarrow}/(2f_{\downarrow}-1)$, respectively. The gauge potentials ${\mathbf A}^{\sigma}$ are the same for all the partons so we omit the superscript. We denote the parton fields as ${\widetilde\psi}_{{\rm I}\sigma}$ and ${\widetilde\psi}_{\rm II}$ to construct the Lagrangian density
\begin{eqnarray}
\mathcal{L}_{\rm parton} &=& \sum_{\sigma=\uparrow,\downarrow} {\widetilde\psi}^{*}_{{\rm I}\sigma}(t,{\mathbf r}) \left[ \left( i\partial_t - a_0 \right)- \frac{1}{2M} \left( {\mathbf p} - e_{{\rm I}\sigma} {\mathbf A} + {\mathbf a} \right)^2 \right] {\widetilde\psi}_{{\rm I}\sigma}(t,{\mathbf r}) \nonumber \\
&+& {\widetilde\psi}^{*}_{\rm II}(t,{\mathbf r}) \left[ \left( i\partial_t + a_0 \right) - \frac{1}{2M} \left( {\mathbf p} - e_{\rm II} {\mathbf A} - {\mathbf a} \right)^2 \right] {\widetilde\psi}_{\rm II}(t,{\mathbf r})
\end{eqnarray}
where the gauge field $a_{\mu}(t,{\mathbf r})$ is used to glue the partons together to satisfy the constraint $\sum_{\sigma=\uparrow,\downarrow} {\widetilde\psi}^{*}_{{\rm I}\sigma}{\widetilde\psi}_{{\rm I}\sigma}={\widetilde\psi}^{*}_{\rm II}{\widetilde\psi}_{\rm II}$. After integrating out the parton fields, we find that the low-energy effective theory for the $\Psi_{[f_{\uparrow},f_{\downarrow}]}$ state is a Chern-Simons theory with action ${\mathcal L}_{\rm eff} = (f_{\uparrow}-f_{\downarrow}+1) \epsilon_{\lambda\mu\nu} a_{\lambda} \partial_{\mu} a_{\nu}/4\pi$. This predicts that the ground state degeneracy of the $\Psi_{[f_{\uparrow},f_{\downarrow}]}$ state on the torus is $f_{\uparrow}-f_{\downarrow}+1$, which agrees with the numerical results for the $f_{\uparrow}=1,f_{\downarrow}=1$ and $f_{\uparrow}=2,f_{\downarrow}=1$ states. The parton theory does not give information about the energetics so it can not tell us that the $f_{\uparrow}=2,f_{\downarrow}=2$ state has a small gap or may even be gapless. For the $f_{\uparrow}=1,f_{\downarrow}=2$ state, the coefficient in the Chern-Simons action vanishes, which suggests that it does not represent a gapped topological phase. One may compute to higher orders of the gauge field $a_{\mu}(t,{\mathbf r})$ and the resulting theory is likely to be a Maxwell theory describing a superfluid state. This is also the case for all other states at $f_{\downarrow}=f_{\uparrow}+1$.

\end{widetext}

\begin{figure}
\includegraphics[width=0.35\textwidth]{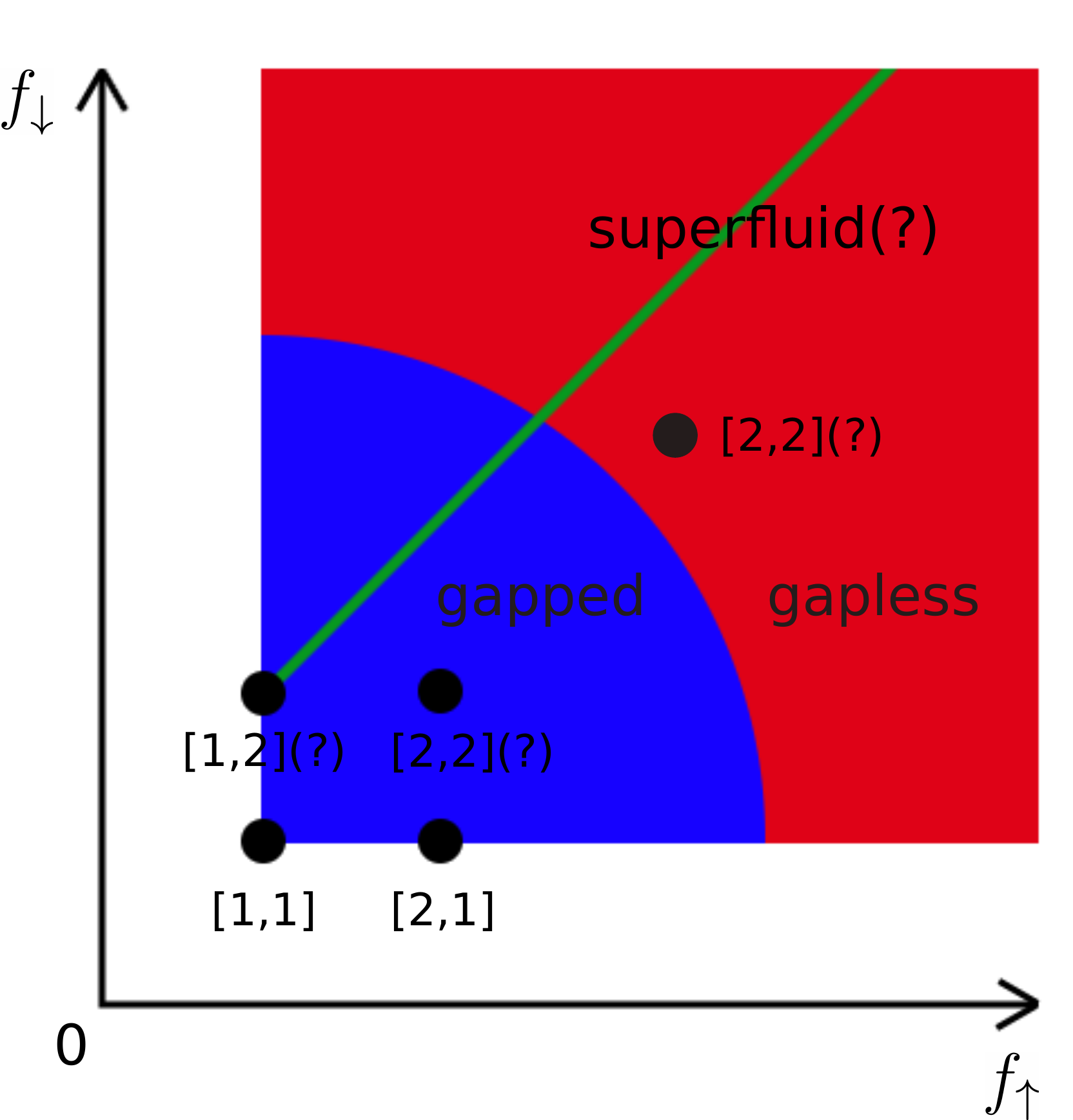}
\caption{A schematic phase diagram of our system. The parameters are given as $[f_{\uparrow},f_{\downarrow}]$ on the figure. The system is gapless for sufficiently large $f_{\uparrow}$ and/or $f_{\downarrow}$ but can be gapped if these two parameters are small enough. The states with $f_{\downarrow}=f_{\uparrow}+1$ may represent a class of exotic superfluid. The $[f_{\uparrow},f_{\downarrow}]=[2,2]$ state is shown twice because it may be gapped or gapless. The question marks are used to indicate that the nature of some states is not fully settled.}
\label{Figure4}
\end{figure}

\section{Conclusion and Discussion}

In conclusion, we have studied possible topological phases of two-component bosons in artificial gauge potentials whose magnitudes for the two components are different. We propose trial wave functions using the composite fermion theory to describe systems with spin-up and spin-down composite fermions moving in effective magnetic fields pointing to opposite directions. An alternative interpretation is given using the parton construction. The trial wave functions are found to be excellent approximations of exact diagonalization results for several systems with contact interaction. An interesting observation is that the low-lying excitations of some states can be modeled by exciting only the spin-up composite fermions. We obtain effective field theories using flux attachment and parton construction, which might seem different but they both predict correct ground state degeneracies as confirmed by numerical calculations. The flux attachment method also gives useful information about the energetics of the states and helps us to understand the numerical results. Our results are summarized in the schematic phase diagram Fig. \ref{Figure4}. The system is gapless for sufficiently large $f_{\uparrow}$ and/or $f_{\downarrow}$ but can be gapped if these two parameters are small enough. It is possible that the $f_{\uparrow}=2,f_{\downarrow}=2$ state is already gapless. The states with $f_{\downarrow}=f_{\uparrow}+1$ may represent a class of exotic superfluid and deserve further investigations.

Our theoretical work points out a direction for the experimental study of artificial gauge potentials. The synthetic magnetic field in a rotating Bose-Einstein condensate depends on the angular velocity, so two internal states will experience different fields if they are rotated at different speeds. It is unclear whether this can be achieved without destabilizing the system. A more probable avenue is to combine two artificial gauge potentials with one being the same for all bosons while the other has opposite directions for the two components. The experimental evidences of the states studied here are similar to those of other quantum Hall states. The compressibility of a cold atom system can be measured to see whether it is gapped \cite{Pethick}. The low-lying excitations can be probed using spectroscopic techniques \cite{Ha}. The gapless edge modes localized at the boundary of a gapped topological state can be visualized in experiments \cite{Goldman2}.

Finally, we would like to suggest a few directions that may be explored in future works. In this paper, we have only studied two-component bosons with contact interaction and the excitations of Eq. \ref{ManyWave} do not possess non-Abelian braid statistics. It is natural to ask what happens if the interaction potential between bosons is long-ranged and how to design the artificial gauge potentials to facilitate the emergence of non-Abelian topological states. One may also study bosonic or fermionic systems with more internal degrees of freedom coupled to general Abelian or non-Abelian artificial gauge potentials.

\section*{Acknowledgement} 

We thank J. K. Jain and J. A. Hutasoit for very helpful discussions and comments on the manuscript. Exact diagonalization calculations are performed using the DiagHam package for which we are grateful to all the authors. This work was supported by the EU integrated project SIQS, the EU integrated project AQUTE, and the US DOE under Grant No. DE-SC0005042. High performance computing resources and services were provided by the Institute for Cyberscience at The Pennsylvania State University.

\begin{appendix}

\begin{widetext}

\section{Hamiltonian Matrix Elements}

For the spherical geometry, the matrix elements are given by
\begin{eqnarray}
F^{\sigma\tau\tau\sigma}_{m_1m_2m_4m_3} = \int d{\mathbf \Omega}_1 d{\mathbf \Omega}_2 \; \left[ \psi^\sigma_{m_1}({\mathbf \Omega}_1) \right]^* \left[ \psi^\tau_{m_2}({\mathbf \Omega}_2) \right]^* \delta({\mathbf r}_1-{\mathbf r}_2) \psi^\tau_{m_4}({\mathbf \Omega}_2) \psi^\sigma_{m_3}({\mathbf \Omega}_1) \nonumber
\end{eqnarray}
where ${\mathbf r}=R(\sin\theta\cos\phi,\sin\theta\sin\phi,\cos\theta)$ and ${\mathbf \Omega}={\mathbf r}/R$. One can show that
\begin{eqnarray}
&& \psi^{N^\sigma_\phi}_{m_1} \psi^{N^\tau_\phi}_{m_2} = (-1)^{N^\sigma_\phi-N^\tau_\phi} S^{N^\sigma_\phi,N^\tau_\phi}_{m_1,m_2} \psi^{N^\sigma_\phi+N^\tau_\phi}_{m_1+m_2} \nonumber \\
&& S^{N^\sigma_\phi,N^\tau_\phi}_{m_1,m_2} = \left[ \frac{(N^\sigma_\phi+1)(N^\tau_\phi+1)}{4\pi(N^\sigma_\phi+N^\tau_\phi+1)} \right]^{1/2} {\rm CG} \nonumber
\end{eqnarray}
where ${\rm CG}=\langle N^\sigma_\phi/2,-m_1;N^\tau_\phi/2,-m_2 | N^\sigma_\phi/2+N^\tau_\phi/2,-m_1-m_2 \rangle$ are Clebsch-Gordon coefficients, so we have
\begin{eqnarray}
F^{\sigma\tau\tau\sigma}_{m_1m_2m_4m_3} &=& \frac{1}{R^2} \int d{\mathbf \Omega}_1 \left[ \psi^{N^\sigma_\phi}_{m_1}({\mathbf \Omega}_1) \right]^* \left[ \psi^{N^\tau_\phi}_{m_2}({\mathbf \Omega}_1) \right]^* \psi^{N^\tau_\phi}_{m_4}({\mathbf \Omega}_1) \psi^{N^\sigma_\phi}_{m_3}({\mathbf \Omega}_1) \nonumber \\
&=& \frac{1}{R^2} S^{N^\sigma_\phi,N^\tau_\phi}_{m_1,m_2} S^{N^\sigma_\phi,N^\tau_\phi}_{m_3,m_4} \int d{\mathbf \Omega}_1 \left[ \psi^{N^\sigma_\phi+N^\tau_\phi}_{m_1+m_2}({\mathbf \Omega}_1) \right]^* \psi^{N^\sigma_\phi+N^\tau_\phi}_{m_3+m_4}({\mathbf \Omega}_1) \nonumber \\
&=& \frac{1}{R^2} \delta_{m_1+m_2,m_3+m_4} S^{N^\sigma_\phi,N^\tau_\phi}_{m_1,m_2} S^{N^\sigma_\phi,N^\tau_\phi}_{m_3,m_4} \nonumber
\end{eqnarray}

For the torus geometry, the matrix elements are given by
\begin{eqnarray}
F^{\sigma\tau\tau\sigma}_{m_1m_2m_4m_3} = \int d^2 {\mathbf r}_1 d^2 {\mathbf r}_2 \; \left[ \psi^\sigma_{m_1}({\mathbf r}_1) \right]^* \left[ \psi^\tau_{m_2}({\mathbf r}_2) \right]^* \delta({\mathbf r}_1-{\mathbf r}_2) \psi^\tau_{m_4}({\mathbf r}_2) \psi^\sigma_{m_3}({\mathbf r}_1) \nonumber
\end{eqnarray}
where ${\mathbf r}=(x,y)$ is the usual two-dimensional coordinates. We define the reciprocal lattice vectors as ${\mathbf G}_1=2\pi{\widehat e}_x/L_1$ and ${\mathbf G}_2=2\pi{\widehat e}_y/L_2$. The interaction potential can be expressed in momentum space as 
\begin{eqnarray}
V({\mathbf r}_1 - {\mathbf r}_2) &=& \frac{1}{L_1L_2} \sum_{\mathbf{q}} e^{i{\mathbf q}\cdot({\mathbf r}_1 - {\mathbf r}_2)} \nonumber
\end{eqnarray}
where ${\mathbf q}=q_1{\mathbf G}_1+q_2{\mathbf G}_2$. One can show that
\begin{eqnarray}
\int d^2 {\mathbf r} \left[ \psi_{m_1}({\mathbf r}) \right]^* e^{i{\mathbf q}\cdot{\mathbf r}} \psi_{m_3}({\mathbf r}) = \exp \left\{ -\frac{1}{4} {\mathbf q}^2 \ell^2_B \right\} \exp \left\{ i\frac{{\pi}q_1}{N_\phi} (m_1+m_3) \right\} {\widetilde\delta}^{N_\phi}_{m_1,m_3+q_2} \nonumber
\end{eqnarray}
so we have
\begin{eqnarray}
F^{\sigma\tau\tau\sigma}_{m_1m_2m_4m_3} = \frac{1}{N^{\uparrow}_{\phi}} \sum_{q_1,q_2} \exp \left\{ -\frac{1}{4} {\mathbf q}^2 (\ell^2_\sigma+\ell^2_\tau) \right\} \sum^{N^\sigma_\phi}_{m_1} \sum^{N^\tau_\phi}_{m_2} \exp \left\{ i2{\pi}q_1 \left[ \frac{(m_1-q_2/2)}{N^\sigma_\phi} - \frac{(m_2+q_2/2)}{N^\tau_\phi} \right] \right\} {\widetilde\delta}^{N^{\rm G}_\phi}_{m_1+m_2,m_3+m_4} \nonumber
\end{eqnarray}
where ${\widetilde\delta}^{N^{\rm G}_\phi}_{i,j}$ and $N^{\rm G}_\phi$ has been defined in the main text.

\section{Electromagnetic Response Function}

In this section, we will use $r$ to denote space-time coordinates ($r_0=t$, $r_1=x$, $r_2=y$) and $q$ to denote frequency-momentum variables ($q_0=\omega$, $q_1=q_x$, $q_2=q_y$). The Chern-Simons gauge field ${\mathbf a}$ can be separated into an average $\langle{\mathbf a}\rangle$ and a fluctuation ${\widetilde{\mathbf a}}^{\sigma}$. The $\langle{\mathbf a}\rangle$ part reduces the gauge field ${\mathbf A}^{\sigma}$ for physical bosons to an effective gauge field ${\mathbf{\overbar A}}^{\sigma}$ for composite fermions. We denote the effective magnetic field strength for composite fermions as ${\overbar B}^{\sigma}$, the effective cyclotron frequency of composite fermions as ${\overbar\omega}_{\sigma}$, and the filling factor of composite fermions as $f^{\sigma}$ (they are the $m$ and $n$ used in the main text). We introduce ${\widetilde{\mathbf b}}={\widetilde{\mathbf a}}-{\mathbf A}^{\sigma}_{\rm ext}$ and $S_{I}=\int d^3 r \; {\mathcal L}_{I}$ with
\begin{eqnarray}
{\mathcal L}_{I} = - {\widetilde b}^{\sigma}_{0} {\widetilde\psi}^{*}_{\sigma}(r) {\widetilde\psi}_{\sigma}(r) -\frac{1}{2M} {\widetilde{\mathbf b}}^{\sigma} \cdot \left[ {\widetilde\psi}^{*}_{\sigma}(r) \left( {\mathbf p} - \mathbf{\overbar A}^\sigma \right) {\widetilde\psi}_{\sigma}(r) - ( {\mathbf p} + \mathbf{\overbar A}^{\sigma}) {\widetilde\psi}^{*}_{\sigma} (r) {\widetilde\psi}_{\sigma}(r) \right] - \frac{1}{2M} {\widetilde{\mathbf b}}^{\sigma} \cdot {\widetilde{\mathbf b}}^{\sigma} {\widetilde\psi}^{*}_{\sigma}(r) {\widetilde\psi}_{\sigma}(r) \nonumber
\end{eqnarray}
The effective action $S_{\rm eff}$ can be computed to the second order of $S_{I}$ as
\begin{eqnarray}
S_{\rm eff} = S_{\rm CS} + \left\langle S_{I} \right\rangle + \frac{i}{2} \left\langle S^{2}_{I} \right\rangle = S_{\rm CS} + \frac{1}{2} \int d^3 r_1 \; d^3 r_2 \; \sum_{\sigma=\uparrow,\downarrow} {\widetilde b}^{\sigma}_{\mu}(r_1) \Pi^{\sigma}_{\mu\nu}(r_1,r_2) {\widetilde b}^{\sigma}_{\nu}(r_2) \nonumber
\end{eqnarray}
where $\Pi_{\mu\nu}$ is the polarization tensor. In terms of the Green's function of the composite fermion field
\begin{eqnarray}
G^{\sigma}_{12} = G^{\sigma}(r_1,r_2) = -i \left\langle {\widetilde\psi}_{\sigma}(r_1) {\widetilde\psi}^{*}_{\sigma}(r_2) \right\rangle \nonumber
\end{eqnarray}
the components of the polarization tensor are
\begin{eqnarray}
\Pi^{\sigma}_{00}(r_1,r_2) &=& i G^{\sigma}_{12} G^{\sigma}_{21} \nonumber \\
\Pi^{\sigma}_{0i}(r_1,r_2) &=& \frac{1}{2M} \Big[ G^{\sigma}_{12} ( \nabla_{2} - i {\mathbf{\overbar A}}^{\sigma} )_i G^{\sigma}_{21} - ( \nabla_2 + i \mathbf{\overbar A}^{\sigma} )_i G^{\sigma}_{12} G^{\sigma}_{21} \Big] \nonumber \\
\Pi^{\sigma}_{i0}(r_1,r_2) &=& \frac{1}{2M} \Big[ ( \nabla_1 - i \mathbf{\overbar A}^{\sigma} )_i G^{\sigma}_{12} G^{\sigma}_{21} - G^{\sigma}_{12} ( \nabla_1 + i \mathbf{\overbar A}^{\sigma} )_i G^{\sigma}_{21} \Big] \nonumber \\
\Pi^{\sigma}_{ij}(r_1,r_2) &=& -\frac{n_{\sigma}}{M} \delta_{ij} \delta(r_1-r_2) - \frac{i}{4M^2} \nonumber \\
&& \times \Big[ ( \nabla_1 - i \mathbf{\overbar A}^{\sigma} )_i G^{\sigma}_{12} ( \nabla_2 - i \mathbf{\overbar A}^{\sigma} )_j G^{\sigma}_{21} + ( \nabla_2 + i \mathbf{\overbar A}^{\sigma} )_j G^{\sigma}_{12} ( \nabla_1 + i \mathbf{\overbar A}^{\sigma} )_i G^{\sigma}_{21} \nonumber \\
&& - ( \nabla_1 - i \mathbf{\overbar A}^{\sigma} )_i ( \nabla_2 + i \mathbf{\overbar A}^{\sigma} )_j G^{\sigma}_{12} G^{\sigma}_{21} - G^{\sigma}_{12} ( \nabla_1 + i \mathbf{\overbar A}^{\sigma} )_i ( \nabla_2 - i \mathbf{\overbar A}^{\sigma} )_j G^{\sigma}_{21} \Big] \nonumber
\end{eqnarray}
where $\nabla_{i}$ ($i=1,2$) takes derivative with respect to ${\mathbf r}_i$.

It is convenient to transform the effective action to momentum space. The first part becomes
\begin{equation}
S_{\rm CS} = \frac{1}{2} \int d^3 q \left[ {\widetilde a}_{0}(-q),{\widehat{\mathbf q}}\cdot{\widetilde{\mathbf a}}(-q),{\widehat{\mathbf q}}\times{\widetilde{\mathbf a}}(-q) \right] C(q) \left[
\begin{array}{c}
{\widetilde a}_{0}(q) \\
{\widehat{\mathbf q}}\cdot{\widetilde{\mathbf a}}(q) \\
{\widehat{\mathbf q}}\times{\widetilde{\mathbf a}}(q)
\end{array}
\right] \nonumber
\end{equation}
where ${\widehat{\mathbf q}}$ is the unit momentum vector, ${\widehat{\mathbf q}}\times{\widetilde{\mathbf a}}(q)$ denotes its $z$ component (the only non-zero component), and the matrix $C(q)$ is
\begin{equation}
\left[
\begin{array}{ccc}
0 & 0 & i\theta |q|  \\
0 & 0 & i\theta q_{0} \\
-i\theta|q| & -i\theta q_{0} & -\theta^{2}|q|^2 V(q)
\end{array}
\right] \nonumber
\end{equation}
with $\theta=(2\pi)^{-1}$ and $V(q)$ being the Fourier transform of the interaction. The second part becomes
\begin{equation}
\frac{1}{2} \int d^3 q \; \sum_{\sigma=\uparrow,\downarrow} {\widetilde b}^{\sigma}_{\mu}(-q) \Pi^{\sigma}_{\mu\nu}(q) {\widetilde b}^{\sigma}_{\nu}(q) \nonumber
\end{equation}
and the components of the polarization tensor are
\begin{eqnarray}
\Pi^{\sigma}_{00}(q) &=& |q|^2 F^{\sigma}_{0}(q) \nonumber \\
\Pi^{\sigma}_{0i}(q) &=& q_0 q_i F^{\sigma}_{0}(q) - i d_\sigma \epsilon_{ij} q_j F^{\sigma}_{1}(q) \nonumber \\
\Pi_{i0}^{\sigma }(q) &=& q_0 q_i F^{\sigma}_{0}(q) + i d_\sigma \epsilon_{ij} q_j F^{\sigma}_{1}(q) \nonumber \\
\Pi_{ij}^{\sigma }(q) &=& \delta_{ij} q^2_0 F^{\sigma}_{0}(q) + i d_\sigma \epsilon_{ij} q_0 F^{\sigma}_{1}(q) + ( \delta_{ij} |q|^2 - q_i q_j ) F^{\sigma}_{2}(q) \nonumber
\end{eqnarray}
where $d_\sigma$ denotes the direction of the effective magnetic field ${\mathbf{\overbar A}}^{\sigma}$. The three quantities $F^{\sigma}_{0,1,2}(q)$ are
\begin{eqnarray}
F^{\sigma}_{0}(q) &=& -\frac{ |{\overbar B}^{\sigma}| e^{-\frac{q^2}{2|{\overbar B}^{\sigma}|}} } {2{\pi}M} \sum^{\infty}_{s=f^{\sigma}} \sum^{f^{\sigma}-1}_{t=0} \frac{s-t}{q^2_0-(s-t)^2{\overbar\omega}^2_{\sigma}} \frac{t!}{s!} \left( \frac{q^2}{2|{\overbar B}^{\sigma}|} \right)^{s-t-1} \left[ L^{s-t}_{t} \left( \frac{q^{2}}{ 2|{\overbar B}^{\sigma}| } \right) \right]^{2} \nonumber \\
F^{\sigma}_{1}(q) &=& \frac{ |{\overbar B}^{\sigma}|^2 e^{-\frac{q^2}{2|{\overbar B}^{\sigma}|}} } {2{\pi}M^2} \sum^{\infty}_{s=f^{\sigma}} \sum^{f^{\sigma}-1}_{t=0} \frac{s-t}{q^2_0-(s-t)^2{\overbar\omega}^2_{\sigma}} \frac{t!}{s!} \left( \frac{q^2}{2|{\overbar B}^{\sigma}|} \right)^{s-t-1} L^{s-t}_{t} \left( \frac{q^{2}}{ 2|{\overbar B}^{\sigma}|} \right)  \nonumber \\
&\times& \left\{ \frac{q^2}{2|{\overbar B}^{\sigma}|} \left[ L^{s-t}_{t} \left( \frac{q^2}{2|{\overbar B}^{\sigma}|} \right) + 2 L^{s-t+1}_{t-1} \left( \frac{q^2}{2|{\overbar B}^{\sigma}|} \right) \left( 1-\delta _{t,0} \right) \right] -(s-t) L^{s-t}_{t}\left(\frac{q^2}{2|{\overbar B}^{\sigma}|} \right) \right\} \nonumber \\
F^{\sigma}_{2}(q) &=&-\frac{ |{\overbar B}^{\sigma}| e^{ -\frac{q^2} {2|{\overbar B}^{\sigma}|} } } {2\pi M^3} \sum^{\infty}_{s=f^{\sigma}} \sum^{f^{\sigma}-1}_{t=0} \frac{s-t}{q^2_0-(s-t)^2{\overbar\omega}^2_{\sigma}} \frac{t!}{s!} \left( \frac{q^2}{2|{\overbar B}^{\sigma}|} \right)^{s-t-1} \nonumber \\
&\times& \left[ L^{s-t}_{t} \left( \frac{q^2}{ 2|{\overbar B}^{\sigma}| } \right) + 2 L^{s-t+1}_{t-1} \left( \frac{q^2}{ 2|{\overbar B}^{\sigma}| } \right) \left( 1-\delta_{t,0} \right) \right] \nonumber \\
&\times& \left\{ \frac{q^2}{ 2 |{\overbar B}^{\sigma}| } \left[ L^{s-t}_{t} \left( \frac{q^{2}}{2|{\overbar B}^{\sigma}|} \right) + 2 L^{s-t+1}_{t-1} \left( \frac{q^2}{2|{\overbar B}^{\sigma}|} \right) \left( 1-\delta_{t,0} \right) \right] - 2(s-t) L^{s-t}_{t} \left( \frac{q^2}{2|{\overbar B}^{\sigma}|} \right) \right\} \nonumber
\end{eqnarray}
where $L^n_m(x)$ is the associated Laguerre polynomial.

We now have the effective action
\begin{eqnarray}
S_{\rm eff} &=& \frac{1}{2}\int d^3 q \left[ {\widetilde a}_{0}(-q),{\widehat{\mathbf q}}\cdot{\widetilde{\mathbf a}}(-q),{\widehat{\mathbf q}}\times{\widetilde{\mathbf a}}(-q) \right] D(q)\left[
\begin{array}{c}
{\widetilde a}_{0}(q) \\
{\widehat{\mathbf q}}\cdot{\widetilde{\mathbf a}}(q) \\
{\widehat{\mathbf q}}\times{\widetilde{\mathbf a}}(q)
\end{array}
\right] \nonumber \\
&-& \sum_{\sigma} \int d^3 q \left[ {\widetilde a}_{0}(-q),{\widehat{\mathbf q}}\cdot{\widetilde{\mathbf a}}(-q),{\widehat{\mathbf q}}\times{\widetilde{\mathbf a}}(-q) \right] E_{\sigma}(q)\left[
\begin{array}{c}
A^{\sigma}_{{\rm ext},0}(q) \\
{\widehat q}\cdot{\mathbf A}^{\sigma}_{\rm ext}(q) \\
{\widehat q}\times{\mathbf A}^{\sigma}_{\rm ext}(q)
\end{array}
\right] \nonumber \\
&+& \frac{1}{2} \sum_{\sigma} \int d^3 q \left[ A^{\sigma}_{{\rm ext},0}(-q),{\widehat q}\cdot{\mathbf A}^{\sigma}_{\rm ext}(-q),{\widehat q}\times{\mathbf A}^{\sigma}_{\rm ext}(-q) \right] E_{\sigma}(q) 
\left[
\begin{array}{c}
A^{\sigma}_{{\rm ext},0}(q) \\
{\widehat q}\cdot{\mathbf A}^{\sigma}_{\rm ext}(q) \\
{\widehat q}\times{\mathbf A}^{\sigma}_{\rm ext}(q)
\end{array}
\right] \nonumber
\end{eqnarray}
where
\begin{equation}
E_{\sigma}(q)=\left[
\begin{array}{ccc}
F^{\sigma}_{0}(q) |q|^{2} & F^{\sigma}_{0}(q) q_{0} |q| & i d_\sigma F^{\sigma}_{1}(q) |q| \\
F^{\sigma}_{0}(q) q_{0} |q| & F_{0}^{\sigma}(q) q^{2}_{0} & i d_\sigma F^{\sigma}_{1}(q) q_{0} \\
-i d_\sigma F^{\sigma}_{1}(q) |q| & -i d_\sigma F_{1}^{\sigma}(q)q_{0} & F^{\sigma}_{0}(q) q^{2}_{0} + F^{\sigma}_{2}(q) |q|^{2}
\end{array}
\right] \nonumber
\end{equation}
and $D(q)=C(q)+\sum_{\sigma=\uparrow,\downarrow} E_{\sigma}(q)$. The inverse of $D(q)$ is
\begin{eqnarray}
D^{-1}(q) &=& \frac{1}{ (q^{2}_{0}+|q|^{2}) \left[ F^{2}_{0}(q)q^{2}_{0} + F_{0}(q)F_{2}(q) |q|^{2} - F_{1}^{2}(q) \right] } \nonumber \\
&\times& \left[
\begin{array}{ccc}
\frac{|q|^{2}}{q^{2}_{0}+|q|^{2}} \left[ F_{0}(q)q^{2}_{0}+F_{2}(q) |q|^{2} \right] & \frac{q_{0}|q|}{q^{2}_{0}+|q|^{2}} \left[ F_{0}(q)q^{2}_{0}+F_{2}(q) |q|^{2} \right]  & -iF_{1}(q)|q|  \\
\frac{q_{0}|q|}{q^{2}_{0}+|q|^{2}} \left[ F_{0}(q)q_{0}^{2}+F_{2}(q)|q|^{2} \right] & \frac{q^{2}_{0}}{q^{2}_{0}+|q|^{2}} \left[ F_{0}(q)q^{2}_{0}+F_{2}(q)|q|^{2} \right]  & -iF_{1}(q) q_{0} \\
iF_{1}(q) |q| & iF_{1}(q)q_{0} & F_{0}(q) \left( q^{2}_{0}+|q|^{2} \right)
\end{array}
\right] \nonumber
\end{eqnarray}
with $F_{0}(q)=\sum_{\sigma=\uparrow,\downarrow} F^{\sigma}_{0}(q)$, $F_{1}(q)=\sum_{\sigma=\uparrow,\downarrow} d_\sigma F^{\sigma=}_{1}(q)+\theta$, and $F_{2}(q)=\sum_{\sigma=\uparrow,\downarrow} F^{\sigma}_{2}(q)-\theta^{2}V(q)$. One can replace the Chern-Simons gauge field $a$ in $\int D[a]$ by its fluctuation ${\widetilde a}$ and the integration gives 
\begin{eqnarray}
S_{\rm ext} = \frac{1}{2} \int d^3 q \sum_{\sigma\tau={\uparrow,\downarrow}} \left[ A^{\sigma}_{{\rm ext},0}(-q),{\widehat q}\cdot{\mathbf A}^{\sigma}_{\rm ext}(-q),{\widehat q}\times{\mathbf A}^{\sigma}_{\rm ext}(-q) \right] K^{\sigma\tau}(q) 
\left[
\begin{array}{c}
A^{\tau}_{{\rm ext},0}(q) \\
{\widehat q}\cdot{\mathbf A}^{\tau}_{\rm ext}(q) \\
{\widehat q}\times{\mathbf A}^{\tau}_{\rm ext}(q)
\end{array}
\right] \nonumber
\end{eqnarray}
where the electromagnetic response function $K^{\sigma\tau}(q) = E_{\sigma}(q) \delta_{\sigma\tau} - E_{\sigma}(q) D^{-1}(q) E_{\tau}(q)$. The dispersion of the collective magnetoplasma modes is determined by the poles of $K^{\sigma\tau}(q)$ at which
\begin{eqnarray}
F^{2}_{0}(q)q^{2}_{0}+F_{0}(q)F_{2}(q)|q|^{2}-F^{2}_{1}(q)=0 \nonumber
\end{eqnarray}
One can solve this equation and get the zero momentum gap
\begin{eqnarray}
\Delta _{0} &=& \frac{1}{2} \left[ \sqrt{(f_{\uparrow}+1)^{2}{\overbar\omega}^{2}_{\uparrow} + (f_{\downarrow}-1)^{2} {\overbar\omega}^{2}_{\downarrow} + 2 (f_{\uparrow}f_{\downarrow}+f_{\uparrow}-f_{\downarrow}+1) {\overbar\omega}_{\uparrow}{\overbar\omega}_{\downarrow}} - (f_{\uparrow}+1){\overbar\omega}_{\uparrow}-(f_{\downarrow}-1){\overbar\omega}_{\downarrow} \right] \nonumber 
\end{eqnarray}

The ground state degeneracy of the many-body states on torus can also be obtained in this approach. In the small $q$ limit, the action becomes
\begin{equation}
S[A_{\rm ext}] = \frac{1}{4\pi} \int d^3 r \; G^{\sigma\tau} \epsilon _{\lambda\mu\nu} A^{\sigma}_{{\rm ext},\lambda}(r) \partial_{\mu} A^{\tau}_{{\rm ext},\nu}(r) \nonumber
\end{equation}
with
\begin{eqnarray}
G^{\sigma\tau} = d_\sigma f^{\sigma} \delta_{\sigma\tau} - \frac{ d_\sigma d_\tau f^{\sigma}f^{\tau} } {\sum_{\sigma} d_\sigma f^{\sigma}+1} \nonumber
\end{eqnarray}
We introduce a gauge field $b$ to express the particle current as $J^{\sigma}_{\lambda}=\epsilon_{\lambda\mu\nu}\partial_{\mu}b^{\sigma}_{\nu}/(2\pi)$ and using the functional bosonization method to obtain the effective action
\begin{eqnarray}
S_{\rm eff} = -\frac{1}{4\pi} \int d^3 r \; Q^{\sigma\tau} \epsilon_{\lambda\mu\nu} b^{\sigma}_{\lambda}(r) \partial_{\mu} b^{\tau}_{\nu}(r) \nonumber
\end{eqnarray}
with $Q=G^{-1}$. The ground state degeneracy of the state on torus is $|{\rm det}Q|$. For the $f^{\uparrow}=1,f^{\downarrow}=1$ state, the $Q$-matrix is 
\begin{eqnarray}
\left(
\begin{array}{cc}
2 & 1 \\
1 & 0
\end{array}
\right) \nonumber
\end{eqnarray}
which predicts one ground state on torus. For the $f^{\uparrow}=2,f^{\downarrow}=1$ state, the $Q$-matrix is
\begin{eqnarray}
\left(
\begin{array}{cc}
3/2 & 1 \\
1 & 0
\end{array}
\right) \nonumber
\end{eqnarray}
However, this matrix does not give a gauge-invariant Chern-Simons theory because $Q^{\uparrow\uparrow}=3/2$ is not an integer. To fix this problem, we introduce an auxilary gauge field $b^{\rm A}_{\mu}$ and enlarge $b$ to $(b^{\rm A}_{\mu},b^{\uparrow}_{\mu},b^{\downarrow}_{\mu})$. The effective action can be rewritten as
\begin{eqnarray}
S_{\rm eff} = -\frac{1}{4\pi} \int d^3 r \; {\widetilde Q}^{\sigma\tau} \epsilon^{\lambda\mu\nu} b^{\sigma}_{\lambda}(r) \partial_\mu b^{\tau}_{\nu}(r) \nonumber
\end{eqnarray}
with
\begin{eqnarray}
{\widetilde Q}=\left(
\begin{array}{ccc}
2 & 1 & 0 \\
1 & 2 & 1 \\
0 & 1 & 0
\end{array}
\right) \nonumber
\end{eqnarray}
which predicts two ground states on torus.

\end{widetext}

\end{appendix}

\end{document}